\documentclass[12pt]{article}
\usepackage[totalheight = 23cm, totalwidth = 17cm]{geometry}
\newcommand{\nin}{\noindent}
\newcommand{\be}{\begin{equation}}
\newcommand{\ee}{\end{equation}}
\newcommand{\bea}{\begin{eqnarray}}
\newcommand{\eea}{\end{eqnarray}}

\newcommand{\nn}{\nonumber\\}

\usepackage{graphicx}

\begin{document}

\begin{center}

{\Large{\bf Inflaton in $R$-dependent potential}}

\vspace{0.3cm}

Jean Alexandre \footnote{jean.alexandre@kcl.ac.uk}

Department of Physics, King's College London, WC2R 2LS, UK

\vspace{0.7cm}
 
{\bf Abstract}

\end{center}

\nin 
We consider a non-minimally coupled inflaton, in a higher order curvature background, leading to a
potential which evolves with the curvature scalar of the Universe, and which describes two regimes. 
The first one is a
de Sitter phase, where the potential is static, and an exact exponential solution is found for the inflaton.
The second regime, triggered by the inflaton reaching a threshold, leads to a power-law expansion, during which the potential
becomes flat, quickly enough for the inflaton never to reach the minimum of the initial symmetry breaking potential. 
This scenario is an
alternative to the inflaton oscillating about a minimum of the potential,
and where preheating is a consequence of the flattening 
of the potential during the power-law expanding phase.

\vspace{0.5cm}

\section{Introduction}

Among the numerous inflationary models, those of type II involve of a spontaneous symmetry breaking
potential, with an inflaton rolling down the concave part of the potential, to end up in one of its a minima \cite{reviews}.
It is known that such concave potentials generate the so-called spinodal instability, where the absence of restoration force
implies that fluctuations are not perturbative.
As long as one neglects perturbations about an equilibrium state, the spinodal instability can be disregarded. 
But if one wishes to take into account quantum effects, such a concave potential is not realistic: 
it is known that the effective potential of a scalar theory is necessarily convex \cite{convex}. 

It was suggested in \cite{kofman} and in \cite{devega} that quantum fluctuations can be sufficient to 
restore symmetry, where this idea was also supported by numerical studies. In these works, spinodal instability
is related to preheating, generated by the growth of quantum fluctuations. We will recover these ideas here,
explaining that the restoration of symmetry is achieved by a flat effective potential.
 
An explicit flattening of the potential is shown in \cite{holman}, where the authors consider a
description based on the spinodal decomposition, using a potential function of two variables: the mean inflaton field 
and its fluctuations condensate, in the framework of a Hartree approximation. In this way, they obtain
a potential which flattens as the condensate increases, taking thus into account 
the effects of quantum fluctuations. This work was also used in \cite{tsujikawa} for the study of particle production.

We propose here another effective description, taking into account the time-dependence of the inflaton potential,
where the latter evolves with the curvature scalar of the Universe,
mimicking a Wilsonian coarse grained potential which evolves with the decreasing energy scale. Indeed, as was shown in \cite{induced},
exact renormalization group transformations in the presence of a spinodal instability lead to a flat effective
potential which interpolates the minima of the bare potential. 
The resulting model is a non-minimally coupled inflaton, in a higher order curvature 
metric, similarly to other works already done \cite{non-minimal}.
The scenario we obtain contains then two phases:
\begin{itemize}
\item Inflation is provided by a de Sitter phase, where we find an exact solution for the inflaton, which
increases exponentially in time as long as it is in the spinodal region. During this inflationary period, the potential
seen by the inflaton is static, since the curvature scalar is constant. We check that the backreaction of 
possible inflaton fluctuations on the metric are negligible;
\item The end of inflation occurs when the inflaton reaches a certain threshold to be defined.
Assuming a power-law expansion, one finds that the scalar curvature decreases quickly enough
for the potential to flatten before the inflaton reaches a minimum of the bare potential. As a consequence, the
naive oscillations around the minimum do not occur, as was initially proposed in \cite{instant},
and we show that this stage can indeed be identified with a preheating period, where the non-perturbative 
feature corresponds to the potential flattening in the spinodal region.
\end{itemize}

The next section starts with a model of non-minimally coupled inflaton, with higher-order curvature
terms. The non-minimal coupling of the inflaton with the metric plays the role of a potential for the inflaton. The 
former is therefore not static in general, since its characteristics depend on the metric.
The Appendix explains some properties of the effective potential, in a Quantum Field Theory, and shows
how the present potential was motivated, although the present study is based on classical Physics.

\section{Inflaton dynamics}

Motivated by the construction of a potential which can evolve with the geometric characteristics of the Universe, 
we assume the following symmetry breaking potential, based on a non-minimally coupled inflaton,
\bea\label{infpot}
\mbox{if}~~\phi^2\le\phi^2_{sp}~~~~U(\phi)&=&\frac{3\xi}{2g}(R-\mu^2)(2m^2-\xi R+\xi \mu^2)-\frac{\xi}{2}(R-\mu^2)\phi^2\\
\mbox{if}~~\phi^2>\phi^2_{sp}~~~~U(\phi)&=&\frac{3m^4}{2g}-\frac{1}{2}m^2\phi^2+\frac{g}{24}\phi^4\nn
\mbox{with}~~~~\phi^2_{sp}&=&\frac{6}{g}(m^2-\xi R+\xi \mu^2),\nonumber
\eea
where $R$ is the curvature scalar of the Universe, and $\mu^2,\xi$ are parameters to be determined later. The field
$\phi_{sp}$ defines the limits of what will be called the spinodal region $\phi^2\le\phi^2_{sp}$. 
The limit $\phi_{sp}$ is smaller than the minimum of the 
symmetry breaking potential, and varies with $R$. In the spinodal region, the non-minimal coupling plays the role of a 
quadratic potential for the inflaton, with a negative slope. 
The higher order curvature term is present in order to have a vanishing minimum for the potential. 
This potential is motivated by results in exact renormalization studies, as explained in the Appendix, and is sketched in fig.\ref{Ueff}. It
leads to the following actions to study ($m_{pl}$ is the Plank mass):
\begin{itemize}
\item Inside the spinodal region, for $\phi^2\le \phi^2_{sp}$:
\bea\label{Sin}
S_{in}&=&\int d^4x\sqrt{-g}\Bigg\lbrace \left( \frac{m_{pl}^2}{16\pi}-\frac{3\xi}{g}(m^2+\xi\mu^2)+
\frac{\xi}{2}\phi^2\right) R\nn
&&~~~~~~~~~~-\frac{1}{2}g_{\mu\nu}\partial^\mu\phi\partial^\nu\phi+\frac{3\xi^2}{2g}R^2
-\frac{\xi}{2}\mu^2\phi^2+\frac{3\xi \mu^2}{2g}(2m^2+\xi \mu^2) \Bigg\rbrace,
\eea
such that the inflaton sees a potential which can depend on time, since this potential varies with $R$;
\item Outside the spinodal region, for $\phi^2>\phi^2_{sp}$:
\be\label{Sout}
S_{out}=\int d^4x\sqrt{-g}\left\lbrace\frac{m_{pl}^2}{16\pi}R-\frac{1}{2}g_{\mu\nu}\partial^\mu\phi\partial^\nu\phi
-\frac{3m^4}{2g}+\frac{1}{2}m^2\phi^2-\frac{g}{24}\phi^4\right\rbrace,
\ee
where the inflaton sees a static potential.
\end{itemize}

\begin{figure}[ht]
\centering
\includegraphics[width=12.6cm]{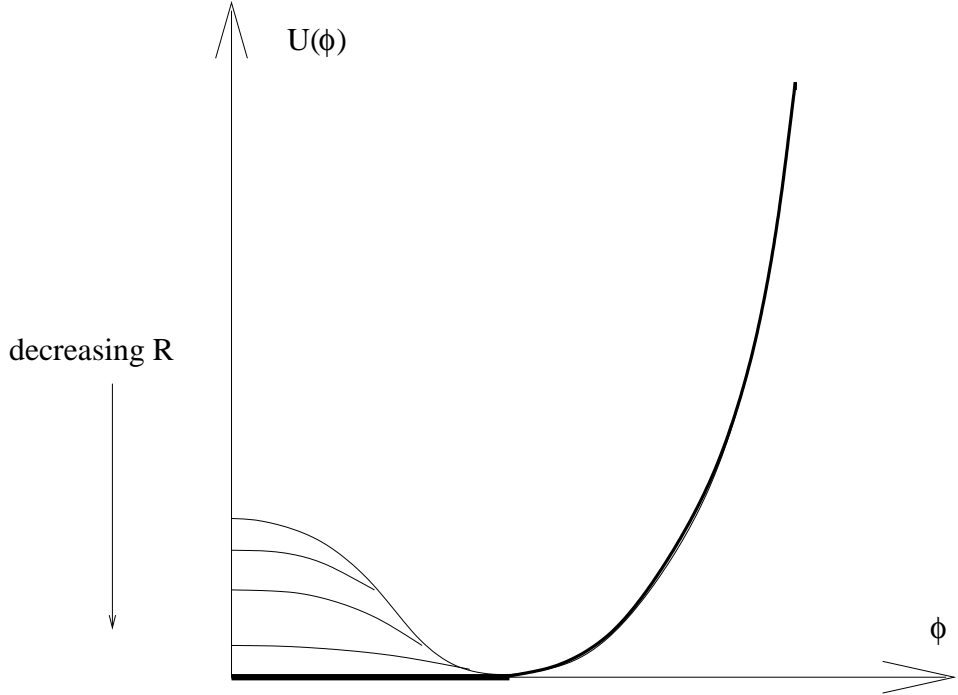}
\caption{The evolution of the inflaton potential with $R\ge\mu^2$. 
The thick line represents the potential for $R=\mu^2$, with the flat
region. This graph is inspired from the results found in \cite{induced}.}
\label{Ueff}
\end{figure}

\subsection{Equations of motion}

From now on, we consider the metric $ds^2=-dt^2+a^2(t)dr^2$, and an inflaton which depends on time only. 
The scalar curvature is then 
\be
R=6\left(\frac{\ddot a}{a}+\frac{\dot a^2}{a^2}\right),
\ee
and a dot represents a time derivative. The actions to study are:
\begin{itemize}
\item in the spinodal region
\bea
S_{in}[a,\phi]&=&\int d^4x~a^3\Bigg\lbrace \frac{\xi}{2}(R-\mu^2)\phi^2
+\left(\frac{m_{pl}^2}{16\pi}-\frac{3\xi}{g}(m^2+\xi \mu^2)\right)R\nn
&&~~~~~~~~~~~~+\frac{3\xi^2}{2g}R^2+\frac{1}{2}\dot\phi^2+\frac{3\xi \mu^2}{2g}(2m^2+\xi \mu^2) \Bigg\rbrace, 
\eea
which leads to the following equations of motion:
\be\label{equamotphiin}
\frac{\delta S_{in}}{\delta\phi}=0=\ddot\phi+3H\dot\phi-\xi(R-\mu^2)\phi,
\ee 
and 
\bea\label{equamotain}
\frac{\delta S_{in}}{\delta a}=&0&=\frac{3\xi \mu^2}{2g}(2m^2+\xi \mu^2)
+2\left(\frac{m_{pl}^2}{16\pi}-\frac{3\xi}{g}(m^2+\xi \mu^2)\right)(2\dot H+3H^2)\\
&&+\frac{1}{2}\dot\phi^2+2\xi\left(\dot\phi^2+\phi\ddot\phi+2H\phi\dot\phi\right)
+\xi\left(2\dot H+3H^2-\frac{1}{2}\mu^2\right)\phi^2\nn
&&+\frac{18\xi^2}{g}\left(2 H^{(3)}+12 H\ddot H+9\dot H^2+18H^2\dot H\right),\nonumber
\eea
where $H^{(3)}$ is the third time derivative of the Hubble parameter $H=\dot a/a$;

\item outside the spinodal region
\be
S_{out}[a,\phi]=\int d^4x~a^3\left\lbrace\frac{m_{pl}^2}{16\pi}R+\frac{1}{2}\dot\phi^2
-\frac{3m^4}{2g}+\frac{1}{2}m^2\phi^2-\frac{g}{24}\phi^4\right\rbrace,
\ee
leading to the equations of motion:
\be\label{equamotphiout}
\frac{\delta S_{out}}{\delta\phi}=0=\ddot\phi+3H\dot\phi-m^2\phi+\frac{g}{6}\phi^3,
\ee 
and
\be\label{equamotaout}
\frac{\delta S_{out}}{\delta a}=0=
\frac{m^2_{pl}}{8\pi}\left(2\dot H+3H^2\right)
+\frac{1}{2}\dot\phi^2-\frac{3m^4}{2g}+\frac{1}{2}m^2\phi^2-\frac{g}{24}\phi^4.
\ee
\end{itemize}
In what follows, we will consider $R>\mu^2$, in order not to have a flat potential from the beginning.

\subsection{De Sitter inflation}

We look here for a solution where $H=H_0=$ constant, such that $R=R_0>\mu^2$ implies $12H_0^2> \mu^2$.
Eq.(\ref{equamotphiin}) gives 
\be\label{equamotphidesitter}
\ddot\phi+3H_0\dot\phi=\xi(12 H_0^2-\mu^2)\phi
\ee
for which solutions are $\phi=\Phi e^{\omega_1 t}$, where 
\be
\omega_1^\pm=\frac{3H_0}{2}\left(-1\pm\sqrt{1+\frac{4\xi}{9}\left(12-\frac{\mu^2}{H_0^2}\right) }\right),
\ee
and $\omega^+_1>0$, whereas $\omega^-_1<0$. We are interested in the increasing solution, and therefore will
disregard $\omega_1^-$. \\
To solve eq.(\ref{equamotain}), we note that the first line of this equation is a contant, that we impose to vanish. This leads 
to the following constraint on the parameters of the model:
\be
\label{condition}
\frac{3\xi \mu^2}{2g}(2m^2+\xi \mu^2)+\frac{3m_{pl}^2}{8\pi}H_0^2=\frac{18\xi}{g}(m^2+\xi \mu^2)H_0^2.
\ee
Eq.(\ref{equamotain}) gives then
\be\label{equamotgdesitter}
\frac{1}{2}\dot\phi^2+2\xi\left(\dot\phi^2+\phi\ddot\phi+2H_0\phi\dot\phi\right)
+\xi\left(3H_0^2-\frac{1}{2}\mu^2\right)\phi^2=0
\ee
for which solutions are $\phi=\Phi e^{\omega_2 t}$, where
\be
\omega_2^\pm=\frac{4\xi H_0}{1+8\xi}\left(-1\pm\sqrt{1-\frac{1+8\xi}{16\xi}\left(6-\frac{\mu^2}{H_0^2}\right) }\right).
\ee
We can see that, if $\mu^2>6H_0^2$, $\omega_2^+$ is positive and can be identified to $\omega_1^+$.
Together with the bound $R_0>\mu^2$, we impose then
\be
6H_0^2<\mu^2<12H_0^2,
\ee
and $\omega_2^+=\omega_1^+$ gives
\be
3(1+8\xi)\left(-1+\sqrt{1+\frac{4\xi}{9}\left(12-\frac{\mu^2}{H_0^2}\right)}\right) 
=8\xi\left(-1+\sqrt{1+\frac{1+8\xi}{16\xi}\left(\frac{\mu^2}{H_0^2}-6\right)}\right),
\ee
which is to be understood as a condition the parameters $\xi,\mu^2,H_0$ have to satisfy. 
For example, if $\mu^2=10H_0^2$, we find numerically 
$\xi\simeq 0.6$. The condition (\ref{condition}) leads then to 
\be\label{assumptionH0m}
10.8~H_0^2-7.2~m^2\simeq \frac{g}{8\pi}m^2_{pl},
\ee
which is a reasonable assumption.\\
We therefore showed here that an exponentially growing inflaton is an exact solution of the equations of motion, 
if we assume a de Sitter inflation, and no slow roll approximation was necessary. 
During this period of inflation, the potential seen by the inflaton is static, since $R=R_0$ is constant.
This regime is valid until the inflaton arrives at the border of the spinodal region, defined by
$\phi^2= \phi_{sp}^2=6(m^2-\xi R_0+\xi \mu^2)/g$, after which it sees the quartic potential
outside the spinodal region, given in eqs.(\ref{infpot}).
The time $t_0$ when inflation stops is then given by
\be
\Phi^2 e^{2\omega^+ t_0}=\frac{6}{g}(m^2-12\xi H_0^2+\xi \mu^2),
\ee
where we note $\omega^+=\omega^+_1=\omega^+_2$. The number of e-foldings the Universe has undertaken is then
\be
N=H_0t_0=\frac{H_0}{2\omega^+}\ln\left(\frac{m^2-12\xi H_0^2+\xi \mu^2}{g\Phi^2/6}\right)
\ee
and can take many values, depending on the parameters of the model, but mainly on the initial amplitude $\Phi$. 
If we consider the previous example, where $\mu^2=10H_0^2$ and thus $\xi\simeq 0.6$, we obtain
\be\label{efoldings}
N\simeq 1.4\times\ln\left( \frac{m^2-1.2H_0^2}{g\Phi^2/6}\right), 
\ee
such that we need $m^2>1.2H_0^2$, which is consistent with the assumption (\ref{assumptionH0m}),
since the latter implies $m^2<1.5H_0^2$. To fix the idea, we will take later $m^2\simeq 1.3H_0^2$.

\subsection{Stability of the de Sitter phase}

The previous subsection is based on classical equations of motion for the inflaton and the metric, and we
study here the stability of the de Sitter metric against possible fluctuations of the inflaton. 
Indeed, gravitational backreaction on a classical de Sitter background can have large effects
(see \cite{backreaction1},\cite{backreaction2}, and references therein, 
where conditions are derived for backreactions to be perturbative, and where studies beyond the first
order of fluctuations are done).\\
We start from the equation of motion (\ref{equamotphidesitter}), and consider its linear perturbation 
arising from the small fluctuations $\delta\phi$ and $\delta H_0$ at a given time. Taking into account the solution 
$\phi=\Phi e^{\omega t}$, where $\omega=\omega_1^+=\omega_2^+$, such that
\be\label{deltaphidots}
\delta\dot\phi=\omega\delta\phi~~~~~~~~~~\delta\ddot\phi=\omega^2\delta\phi,
\ee
we obtain then
\be\label{fluctuations}
\frac{\delta H_0}{H_0}=\frac{\delta\phi}{\phi}~\frac{\xi(12H_0^2-\mu^2)-\omega^2-3\omega H_0}{3H_0(\omega-8\xi H_0)}.
\ee
Note that the second equation of motion (\ref{equamotgdesitter}) does not give any more information, 
since it is quadratic in $\phi$ and, after differentiation and using eqs.(\ref{deltaphidots}), a division by $\phi$ leads to 
the relation (\ref{fluctuations}), provided $\xi\simeq 0.6$, where the two equations of motion are consistent.
If, for this value of $\xi$, we take $\mu^2=10H_0^2$, we obtain then $\omega\simeq0.357H_0$, 
and the perturbations are related by
\be\label{perturb1}
\left|\frac{\delta H_0}{H_0}\right|=\epsilon_1\left|\frac{\delta\phi}{\phi}\right|~~~~~~~~
\mbox{where}~\epsilon_1\simeq 10^{-4}.
\ee
The action (\ref{Sin}) is quadratic in the inflaton, such that one can assume that the spectral modes of $\delta\phi$ 
are uncorrelated and lead to the diffusion law $\left|\delta\phi\right|\simeq H_0\sqrt{H_0t}$ \cite{backreaction2}.
We obtain then, from eq.(\ref{perturb1}),
\be\label{perturb2}
\left|\frac{\delta H_0}{H_0}\right|\simeq \epsilon_1\frac{H_0}{\Phi}\sqrt{H_0t}~e^{-\omega t},
\ee
and if we use the number of e-foldings (\ref{efoldings}), with $m^2=1.3H_0^2$ as stated above, 
this ratio is, at the end of inflation,
\be
\left|\frac{\delta H_0}{H_0}\right|_{end}=\epsilon_2\sqrt{gN}~\exp(\epsilon_3N),
~~\mbox{where}~~\epsilon_2\simeq 1.3\times\epsilon_1 ~~\mbox{and}~~\epsilon_3\simeq 1.4\times10^{-4},
\ee
which is very small for a typical value like $N=70$. It is therefore consistent to assume that backreactions 
of inflaton fluctuations on the metric can be neglected in this model. Note that this result is valid because:
\begin{itemize}
\item The non-minimal coupling in the action (\ref{Sin}) gives rise to the term $\xi(12H_0^2-\mu^2)$ in the numerator of
eq.(\ref{fluctuations}), which almost compensates the derivatives terms $\omega^2+3\omega H_0$, and leads to 
the small parameter $\epsilon_1$ and therefore $\epsilon_2$;
\item The inflaton grows exponentially, and, at the end of inflation,
the small factor $e^{-\omega t_0}\simeq e^{-0.357N}$ in eq.(\ref{perturb2}) almost compensates the large factor 
$H_0/\Phi\propto \sqrt{ge^{N/1.4}}$, which leads to the small parameter $\epsilon_3$.
\end{itemize}
These two reasons are actually not independent, since the specific exponential solution for the inflaton is 
a result of the non-minimal coupling term in the original action.

\subsection{Preheating}

As stated before, inflation ends when the dilaton $\phi$ reaches the limit of the spinodal region $\phi_{sp}$, since the 
dilaton sees then another potential, given in action (\ref{Sout}). The potential is continuous at the junction 
between the spinodal region and
the outside, but differentiable once only, such that its second derivative, related to the effective mass of the 
dilaton, is discontinuous at $\phi=\phi_{sp}$. 
The transition between inflation and preheating is a consequence of this non-differentiable phenomenological
potential.

After exiting the spinodal region, 
the inflaton naively rolls down the minimum of the bare potential and oscillates about the minimum. 
This picture would be correct if the potential was static, but if the Universe enters a power-law expanding regime, the 
curvature scalar $R$ decreases and the potential flatens, in the spinodal region. We will see that this flattening 
occurs much quicker than the inflaton needs to reach the minimum of the bare potential, such that the inflaton
actually does not oscillates. A similar situation happens in the preheating scenario of \cite{instant}, and we will
see that this non-perturbative potential flattening can indeed correspond to a preheating period.\\
To start with, let us review the naive oscillation mechanism.
We assume small oscillations about the minimum of the bare potential, occurring at $\phi_{min}$ given by
$\phi^2_{min}=6m^2/g$. The angular frequency of the oscillations is $\sqrt 2 m$, and we look for a solution of the form
\be\label{oscillate}
\phi\simeq\phi_{min}+At^q\cos(\sqrt 2mt),
\ee 
where $At^q<<\phi_{min}$. Assuming a power-law expanding Universe, with $H=p/t$, it is easy to see that
the equations of motion (\ref{equamotphiout}) and (\ref{equamotaout}) are satisfied to the first order 
in $At^q/\phi_{min}$ if
\be
p=\frac{2}{3}~~~~~~~~\mbox{and}~~~~~~~~q=-1.
\ee
Let us come back now to the evolution of the potential inside the spinodal region.
With the solution $p=2/3$, the curvature scalar is 
\be
R=\frac{4}{3t^2},
\ee
and decreases from $R_0=4/(3t_0^2)$ until it reaches the value $\mu^2$. This takes the time 
\be
\Delta t=\frac{2}{\sqrt 3}\left( \frac{1}{\mu}-\frac{1}{\sqrt R_0}\right),
\ee
that we have to compare with the period of oscillations $\tau=2\pi/(\sqrt 2m)$ of the solution (\ref{oscillate}).
For this, we come back to the example where $\mu^2=10 H_0^2$, and we consider the case $m^2=1.3H_0^2$, as
suggested earlier. We find then
\be
\Delta t\sim 0.01 \tau,
\ee
which means that the inflaton does not have time to oscillate: almost as soon as it exits the spinodal region, 
the potential flatens, which corresponds to a non-perturbative mechanism. \\
We can now estimate the energy lost by the inflaton during this phase. 
At the end of inflation, the inflaton is at the border of the spinodal instability $\phi=\phi_{sp}$,
such that the potential is
\be
U(\phi_{sp})=\frac{3\xi^2}{2g}(R_0-\mu^2)^2,
\ee
and it vanishes at the time $t_0+\Delta t$.
With our previous example $\mu^2=10H_0^2$, we find that the potential energy loss is then
\be\label{loss}
U(\phi_{sp})\simeq\frac{2.1}{g}H_0^4,
\ee
and, assuming an instant thermalization, the corresponding temperature is then
\be
T\sim g^{-1/4}H_0,
\ee 
which can be sufficient to reach a GUT temperature $\sim 10^{-3}m_{pl}$, if we assume the expected orders of 
magnitude $g\sim 10^{-12}$ and $H_0\sim 10^{-6}m_{pl}$.

\section{Concluding remarks}

As the potential inside the spinodal region becomes flat, quantum fluctuations of the inflaton become 
large, and can take any value between the minima of the initial potential. But the slope of the potential
outside the spinodal region generates restoring forces, imposing oscillations of the inflaton 
inside the spinodal region. These large oscillations lead to the decay of the inflaton. 
As a consequence, the Universe is
filled by the decay products resulting from this preheating period, and equations of motion
involving the inflaton only are not valid anymore.
What is happening after depends on these particles produced, and on their coupling to
gravity, but this is left for a future work.

We note that the spinodal instability, leading to a convex potential, is 
universal and independent of the precise shape of the potential. The present study was done with a quartic potential, 
but a cosine, as in the natural inflation scenario \cite{natural}, would lead to a similar mechanism.

In the present work, backreaction of particles produced has been neglected, but taking these into account
would not change the inflationary period described here, as well as the exiting of the spinodal region. The preheating
period, though, might see its metric modified by the decay products, and this is a possible extension of this work.

Finally, we note that, when $R$ reaches $\mu^2$, the potential is flat and the Universe is filled with products
of the inflaton decay, which, unlike the previous inflaton, are expected to be minimally coupled to gravity, 
and another phase of the Universe begins.

\vspace{1cm}

\nin{\bf Acknowledgements} This work is partially supported by the Royal Society, UK.

\section*{Appendix}

We review here some properties of the effective potential in a quantum field theory, leading to what motivated the 
model (\ref{infpot}).
We first show that the dressed potential is necessarily convex. Then
we show the equivalence of the effective potential defined by the Legendre transform, which is the derivative-independent
part of the proper graph generator functional, and the effective potential defined in the Wilsonian sens. 
Finally, we show how the coarse grained potential, defined
by Wilsonian renormalization, interpolates the bare ``Mexican-hat'' potential with the dressed flat potential. 
This interpolation is obtained when the energy scale decreases, and the flat potential is obtained in the IR limit
of the Wilsonian renormalization group transformations.

\subsection{Convexity}

We review here the construction of the effective potential, which contains the interactions dressed by quantum fluctuations,
and we show that it is necessarily convex \cite{convex}. \\ 
Since we are interested in the effective potential rather than the full effective action, 
we will consider a constant classical field and therefore define the
partition function for a constant source $j$, which is, for a Euclidean metric,
\be
Z(j)=\int{\cal D}[\tilde\phi]\exp\left(-S[\tilde\phi]-{\cal V}j\tilde\Phi\right) ,
\ee
where ${\cal V}$ is the volume of space time, and $\tilde\Phi$ is the constant mode of the dynamical variable $\tilde\phi$.
The connected graphs generator functional is $W(j)=-\ln Z(j)$ and has derivative
\be\label{classphi}
\frac{dW}{dj}=\frac{{\cal V}}{Z}\int{\cal D}[\tilde\phi]~\tilde\Phi~\exp\left(-S[\tilde\phi]-{\cal V}j\tilde\Phi\right)
={\cal V}\left<\tilde\Phi\right>\equiv {\cal V}\Phi,
\ee
where $\Phi$ defines the corresponding constant classical field. It is easy to see that the second derivative of $W$ is
\be
W^{(2)}\equiv\frac{d^2 W}{dj^2}={\cal V}^2\left(\left<\tilde\Phi\right>^2-\left<\tilde\Phi^2\right>\right)~<0,
\ee
and is negative, for any value of the source $j$.
The effective potential, function of the classical field, is defined as the Legendre transform of $W$ 
\be
U_{eff}(\Phi)=\frac{1}{{\cal V}}W(j)-j\Phi,
\ee
where the source has to be seen as a functional of the classical field, by inverting the relation (\ref{classphi}).
The first derivative of the effective potential is 
\be
\frac{dU_{eff}}{d\Phi}=\frac{1}{{\cal V}}\frac{d W}{dj} \frac{dj}{d\Phi}-\frac{dj}{d\Phi}\Phi-j=-j,
\ee
and the second derivative is then 
\be
\frac{d^2U_{eff}}{d\Phi^2} =-\frac{dj}{d\Phi}=-\left(\frac{d\Phi}{dj}\right)^{-1}=-\frac{{\cal V}}{W^{(2)}}>0,
\ee
and is positive, for any value of the field $\Phi$. Therefore, the effective potential of the theory must be convex: 
even if we start with a spontaneous symmetry potential with a concave part, when dressing the system, quantum fluctuations
rub out the classical spontaneous symmetry breaking feature and lead to a convex effective potential.

\subsection{Equivalence with the Wilsonian effective potential}

(I would like to thank Janos Polonyi for the argument given here.)\\
The effective potential defined previously, as the Legendre transform of $W$, is the non-derivative part of the 
proper graphs generator functional, and is denoted $U_{prop}$. We show here that it is equivalent to 
the Wilsonian effective potential, denoted $U_{Wils}$. 
We note that for this argument to be valid, it is essential that we work in Minkowski space time, 
so as to be able to express the Dirac distribution in eq.(\ref{delta}) in terms of its Fourier transform.\\
For a constant IR configuration $\Phi$, the Wilsonian effective potential is defined by
\be
\exp\left( i{\cal V}U_{Wils}(\Phi)\right) =
\int{\cal D}[\phi]\exp\left( iS[\Phi+\phi]\right),
\ee
where $S$ is the  bare action defined at some cut off $\Lambda$, and the
dynamical variable $\phi$ which is integrated out has non-vanishing Fourier components for $|p|\le\Lambda$.
One can also write the previous definition as
\bea\label{delta}
\exp\left( i{\cal V}U_{Wils}(\Phi)\right) &=&
\int{\cal D}[\phi]\exp\left( iS[\phi]\right)\delta\left( \int_x\phi-{\cal V}\Phi\right) \\
&=&\int{\cal D}[\phi]\exp\left( iS[\phi]\right)\int_j\exp\left\{ij\left(\int_x\phi-{\cal V}\Phi\right)\right\},\nonumber
\eea
where $\int_x$ denotes the integration over space time,
$j$ is a real variable, and $\int_j$ denotes the integration over $j$. This leads to 
\bea
\exp\left( i{\cal V}U_{Wils}(\Phi)\right) &=&
\int_jZ(j)\exp\left( -i\int_x j\Phi\right) \nn
&=&\int_j \exp\left(iW(j)-i\int_x j\Phi\right)\nn
&=& \int_j \exp\left(i{\cal V}U_{prop}(\Phi)\right).
\eea
In the last expression, the integration over $j$ leads to a multiplicative
constant, as $\Phi$ is fixed. Disregarding the $\Phi$-independent terms, we then obtain 
\be
U_{Wils}(\Phi)=U_{prop}(\Phi),
\ee
which shows the equivalence between the two effective potentials.

\subsection{Towards a flat potential}

Since both definitions of the effective potential are equivalent, Wilsonian transformations towards the IR, 
which describe how a system gets dressed by quantum fluctuations, should build
a convex potential from a symmetry breaking bare potential. It is indeed the case, and we explain here
the mechanism.\\
The earliest version of exact renormalization group equations was derived in \cite{WH}, where a sharp
cut off is used in order to eliminate degrees of freedom. In this work, an effective theory described by the action
$S_{k-\delta k}$, for degrees of freedom
with momenta $|p|\le k-\delta k$, is obtained from a theory described by $S_k$, for degrees of freedom
with momenta $|p|\le k$, by integrating out degrees of freedom with momenta $k-\delta k<|p|\le k$.  
An exact evolution equation for the coarse grained action $S_k$ with the energy scale 
$k$ is then obtain in the limit $\delta k\to 0$. In the situation of a symmetry breaking bare potential,
integrating out degrees of freedom becomes problematic when fluctuations with typical momentum $k$, 
and above the background $\phi$, satisfy
\be
k^2+U_k''(\phi)=0,
\ee
where $U_k$ is the coarse grained potential defined at the scale $k$, and a prime denotes
a derivative with respect to $\phi$.  
This is the spinodal instability, corresponding to the cancellation of restoration force in the system: 
the propagator of fluctuations with momentum $k$ and above the background $\phi$ cannot be defined. 
Classically, the inverse propagator
$k^2+U_{bare}''(\phi)$ can then be negative when $k$ decreases, leading to an ill-defined path integral.\\
An extension of this approach has shown how the Wilsonian flow copes with this spinodal instability \cite{induced},
where it was realized that, when the spinodal instability occurs, non-trivial saddles points appear in the 
functional integral defining the elimination of degrees of freedom. The consequence is that $k^2+U_k''(\phi)$ 
actually remains frozen to 0, for any value of $k$ down to
the IR limit $k=0$. The effective potential $U_{eff}=U_{k=0}$ is then flat since $U_{k=0}''(\phi)=0$.
This mechanism corresponds then to a ``tree-level renormalization'' \cite{tree}: it is similar to 
the famous Maxwell construction in Thermodynamics,
and is also consistent with other works using a Wilsonian approach \cite{Maxwell}. An important point
is that the flat effective potential interpolates the minima of the bare potential, such that the spinodal region 
goes beyond the range of values of $\phi$ for which $U_{bare}''(\phi)<0$. More precisely, 
if we start with the bare potential defined at some cut off $\Lambda$ 
\be\label{quartic}
U_\Lambda(\phi)=\frac{3m^4}{2g}-\frac{m^2}{2}\phi^2+\frac{g}{24}\phi^4,
\ee
where $m^2>0$, it was found numerically in \cite{induced} that 
the coarse grained potential which matches the bare potential outside
the spinodal region is quadratic: 
\be\label{stable}
U_k(\phi)=\frac{3k^2}{2g}(2m^2-k^2)-\frac{k^2}{2}\phi^2~~~~~~~~\mbox{if}~~~~\phi^2\le\frac{6}{g}(m^2-k^2),
\ee
and its evolution with $k$ is sketched in fig.\ref{Ueff}, with the replacement 
\be
k^2\rightarrow\xi(R-\mu^2)
\ee
Note that, for the sake of clarity, we neglect loop corrections to the potential outside the spinodal region.\\
To conclude, as the energy scale $k$ decreases, the potential describing
the system at this scale $k$ gradually rubs out the symmetry breaking features, and in the IR limit becomes completely
flat between the minima of the UV potential.

\end{document}